\documentclass[showpacs,aps]{revtex4}

\usepackage{amsmath,amssymb}
\usepackage{amsthm}
\usepackage{graphics,epsfig,psfrag}

\begin{document}

\theoremstyle{plain}\newtheorem{theorem}{Theorem} 
\theoremstyle{plain}\newtheorem{proposition}{Proposition} 

\title{A Morse-theoretical analysis of gravitational lensing \\
by a Kerr-Newman black hole}
\author{Wolfgang Hasse}
\affiliation{TU Berlin, Sekr.\ PN 7-1, 10623 Berlin, Germany. \\
Wilhelm Foerster Observatory, Munsterdamm 90, 12169 Berlin, Germany. 
Email: astrometrie@gmx.de} 
\author{Volker Perlick}
\affiliation{TU Berlin, Sekr. PN 7-1, 10623 Berlin, Germany. \\
Email: vper0433@itp.physik.tu-berlin.de}


\begin{abstract}
  Consider, in the domain of outer communication $M_+$
  of a Kerr-Newman black hole, a point $p$ (observation  event) and a 
  timelike curve $\gamma$ (worldline of light source). Assume that
  $\gamma$ (i) has no past end-point, (ii) does not intersect the caustic 
  of the past light-cone of $p$, and (iii) goes neither to the horizon 
  nor to infinity in the past. We prove that then for infinitely
  many positive integers $k$ there is a past-pointing lightlike geodesic 
  $\lambda _k$ of (Morse) index $k$ from $p$ to $\gamma$, hence an observer 
  at $p$ sees infinitely many images of $\gamma$. Moreover, we demonstrate 
  that all lightlike geodesics from an event to a timelike curve in $M_+$
  are confined to a certain spherical shell. Our characterization of this 
  spherical shell shows that in the Kerr-Newman spacetime the occurrence 
  of infinitely many images is intimately related to the occurrence of 
  centrifugal-plus-Coriolis force reversal.
\end{abstract}

\pacs{04.70.Bw}

\maketitle


\section{Introduction}\label{sec:intro}
  The question of how many images an observer at an event $p$ sees of a 
  light source with worldline $\gamma$ is equivalent to the question of how 
  many past-pointing lightlike geodesics from $p$ to $\gamma$ exist. In 
  spacetimes with many symmetries this question can be addressed, in principle, 
  by directly integrating the geodesic equation. In the spacetime around a
  non-rotating and uncharged black hole of mass $m$, e.g., which is described
  by the Schwarzschild metric, all lightlike geodesics can be explicitly written  
  in terms of elliptic integrals; with the help of these explicit expressions, 
  it is easy to verify that in the region outside the horizon, i.e. in the 
  region wher $r>2 \, m$, there are infinitely many past-pointing lightlike 
  geodesics from any event $p$ to any integral curve of the Killing vector 
  field $\partial _t$. This was demonstrated already in 1959 bei Darwin 
  \cite{Darwin1959}. We may thus say that a Schwarzschild black hole acts
  as a gravitational lens that produces infinitely many images of any static
  light source. However, already in the Schwarzschild spacetime the problem 
  becomes more difficult if we want to consider light sources which are not
  static, i.e., worldlines $\gamma$ which are not integral curves 
  of $\partial _t$. 

  In this paper we want to investigate this problem for the more general case of a 
  charged and rotating black hole, which is described by the Kerr-Newman metric.
  More precisely, we want to demonstrate that in the domain of outer communication 
  around a Kerr-Newman black hole, i.e., in the domain outside of the outer horizon,
  there are infinitely many past-pointing lightlike geodesics from an unspecified 
  event $p$ to an unspecified worldline $\gamma$, with as little restrictions on 
  $\gamma$ as possible. Although the geodesic equation in the Kerr-Newman spacetime 
  is completely integrable, the mathematical expressions are so involved that it 
  is very difficult to achieve this goal by explicitly integrating the geodesic 
  equation. Therefore it is recommendable to use more indirect methods.

  Such a method is provided by Morse theory. Quite generally, Morse theory relates
  the number of solutions to a variational principle to the topology of the space
  of trial maps. Here we refer to a special variant of Morse theory, developed by 
  Uhlenbeck \cite{Uhlenbeck1975}, which is based on a version of Fermat's principle
  for a globally hyperbolic Lorentzian manifold $(M,g)$. The trial maps are the 
  lightlike curves joining a point $p$ and a timelike curve $\gamma$ in $M$, 
  and the solution curves of Fermat's principle are the lightlike geodesics.
  If $(M,g)$ and $\gamma$ satisfy additional conditions, the topology of the 
  space of trial maps is determined by the topology of $M$. Uhlenbeck's work 
  gives criteria that guarantee the existence 
  of infinitely many past- or future-pointing lightlike geodesics from $p$ to 
  $\gamma$. In this paper we will apply her results to the domain of outer 
  communication around a Kerr-Newman black hole which is, indeed, a globally 
  hyperbolic Lorentzian manifold. 
  We will show that the criteria for having infinitely many past-pointing timelike 
  geodesics from $p$ to $\gamma$ are satisfied for every event $p$ and every timelike 
  curve $\gamma$ in this region, provided that the following three conditions are
  satisfied. First, $\gamma$ must not have a past end-point; it is obvious
  that we need a condition of this kind because otherwise it would be possible
  to choose for $\gamma$ an arbitrarily short section of a worldline such 
  that trivially the number of past-pointing lightlike geodesics from $p$ to
  $\gamma$ is zero. Second, $\gamma$ must not intersect the caustic of the
  past light-cone of $p$; this excludes all cases where $p$ sees an extended
  image, such as an Einstein ring, of $\gamma$. Third, in the past the worldline
  $\gamma$ must not go to the horizon or to infinity. Under these (very mild)
  restrictions on the motion of the light source we will see that the Kerr-Newman black
  hole acts as a gravitational lens that produces infinitely many images. Moreover,
  we will also show that all (past-directed) lightlike geodesics from $p$ to $\gamma$ are confined
  to a certain spherical shell. For the characterization of this shell we will have
  to discuss a light-convexity property which turns out to be intimately related to
  the phenomenon of centrifugal(-plus-Coriolis) force reversal. This phenomenon
  has been discussed, first in spherically symmetric static and then in more 
  general spacetimes, in several papers by Marek Abramowicz with various coauthors;
  material which is of interest to us can be found, in particular, in Abramowicz,
  Carter and Lasota \cite{AbramowiczCarterLasota1988}, Abramowicz \cite{Abramowicz1990} 
  and Abramowicz, Nurowski and Wex \cite{AbramowiczNurowskiWex1993}. 

  The paper is organized as follows. In Section \ref{sec:morse} we
  summarize the Morse-theoretical results we want to use. Section \ref{sec:centrifugal} 
  is devoted to the notions of centrifugal and Coriolis force in the Kerr-Newman 
  spacetime; in particular, we introduce a potential $\Psi _+$ (respectively $\Psi_-$) 
  that characterizes the sum of centrifugal and Coriolis force with respect to 
  co-rotating (respectively counter-rotating) observers whose velocity approaches 
  the velocity of light. In Section \ref{sec:imaging} we discuss multiple imaging 
  in the Kerr-Newman spacetime with the help of the Morse theoretical result quoted 
  in Section \ref{sec:morse} and with the help of the potential $\Psi _{\pm}$
  introduced in Section \ref{sec:centrifugal}. Our results are summarized and
  discussed in Section \ref{sec:conclusion}.

\section{A result from Morse theory}\label{sec:morse}
  In this section we briefly review a Morse-theoretical result that relates the 
  number of lightlike geodesics between a point $p$ and a timelike curve $\gamma$ 
  in a globally hyperbolic Lorentzian manifold to the topology of this manifold. 
  This result was found by Uhlenbeck \cite{Uhlenbeck1975} and its relevance in view of 
  gravitational lensing was discussed by McKenzie \cite{McKenzie1985}. Uhlenbeck's work 
  is based on a variational principle for lightlike geodesics (``Fermat principle'')
  in a globally hyperbolic Lorentzian manifold, and her main method of proof is to 
  approximate trial paths by broken geodesics. With the help of infinite-dimensional
  Hilbert manifold techniques Giannoni, Masiello, and Piccione were able to rederive
  Uhlenbeck's result \cite{GiannoniMasiello1996} and to generalize it to certain 
  subsets-with-boundary of spacetimes that need not be globally hyperbolic
  \cite{GiannoniMasielloPiccione1998}. In contrast to Uhlenbeck, they start out 
  from a variational principle for lightlike geodesics that is not restricted to 
  globally hyperbolic spacetimes. (Such a Fermat principle for arbitrary 
  general-relativistic spacetimes was first formulated by Kovner \cite{Kovner1990}; 
  the proof that the solution curves of Kovner's variational principle are, indeed, 
  precisely the lightlike geodesics was given by Perlick \cite{Perlick1990b}). 
  Although for our purpose the original Uhlenbeck result is sufficient, readers 
  who are interested in technical details are encouraged to also consult the 
  papers by Giannoni, Masiello, and Piccione, in particular because in the 
  Uhlenbeck paper some of the proofs are not worked out in full detail. 
  
  Following Uhlenbeck \cite{Uhlenbeck1975}, we consider a 4-dimensional Lorentzian manifold
  $(M,g)$ that admits a foliation into smooth Cauchy surfaces, i.e., a globally 
  hyperbolic spacetime. (For background material on globally hyperbolic spacetimes 
  the reader may consult, e.g., Hawking and Ellis \cite{HawkingEllis1973}. The fact that the 
  original definition of global hyperbolicity is equivalent to the existence of a 
  foliation into \emph{smooth} Cauchy surfaces was completely proven only recently by 
  Bernal and S{\'a}nchez \cite{BernalSanchez2005}.) Then $M$ can be written 
  as a product of a 3-dimensional manifold $S$, which serves as the prototype for 
  each Cauchy surface, and a time-axis,
\begin{equation}\label{eq:product}
  M= S \times {\mathbb{R}} \, .
\end{equation}
  Moreover, this product can be chosen such that the metric $g$ orthogonally splits 
  into a spatial and a temporal part,
\begin{equation}\label{eq:globhyp}
  g =  g_{ij}(x, t ) \, dx^i \, dx^j - f(x, t ) \, d t ^2 \, ,
\end{equation}
  where $t$ is the time coordinate given by projecting from $M= S \times {\mathbb{R}}$ 
  onto the second factor, $x = (x^1 , x^2 , x^3 )$ are coordinates on $S$, and the 
  summation convention is used for latin indices running from 1 to 3. (We write 
  (\ref{eq:globhyp}) in terms of coordinates for notational convenience only. We do
  not want to presuppose that $S$ can be covered by a single coordinate system.) 
  We interpret the direction of increasing $t$ as the future-direction on $M$. 
  Again following Uhlenbeck \cite{Uhlenbeck1975}, we say that the splitting (\ref{eq:globhyp}) 
  satisfies the {\em metric growth condition\/} if for every compact subset of $S$ 
  there is a function $F$ with
\begin{equation}\label{eq:F}
  \int_{- \infty} ^0 \frac{d t}{F(t)} = \infty
\end{equation}  
  such that for $t \le 0$ the inequality
\begin{equation}\label{eq:growth}
  g_{ij} (x, t ) \, v^i \, v^j \le f(x, t ) \, F( t )^2 \, G_{ij} (x) \, v^i \, v^j
\end{equation}
  holds for all $x$ in the compact subset and for all $(v^1, v^2, v^3)
  \in {\mathbb{R}}^3$, with a time-independent Riemannian metric $G_{ij}$ on 
  $S$. It is easy to check that the metric growth condition assures 
  that for every (smooth) curve $\alpha : [a,b] \longrightarrow S$ there is a
  function $T : [a,b] \longrightarrow {\mathbb{R}}$ with $T (a)=0$ such 
  that the curve $\lambda : [a,b] \longrightarrow M = S \times {\mathbb{R}}, s \longmapsto 
  \lambda (s) = \big( \alpha (s) , T (s) \big)$ is past-pointing and lightlike.
  In particular, the metric growth condition assures that from each point $p$ in $M$
  we can find a past-pointing lightlike curve to every timelike curve that is vertical
  with respect to the orthogonal splitting chosen. In this sense, the metric growth
  condition prohibits the existence of {\em particle horizons}, cf. Uhlenbeck \cite{Uhlenbeck1975} 
  and McKenzie \cite{McKenzie1985}. Please note that our formulation of the metric growth 
  condition is the same as McKenzie's which differs from Uhlenbeck's by interchanging 
  future and past (i.e., $t \longmapsto - t$). The reason is that Uhlenbeck in 
  her paper characterizes {\em future-pointing\/} lightlike geodesics from a 
  point to a timelike curve whereas we, in view of gravitational lensing, 
  are interested in {\em past-pointing\/} ones.  

  For formulating Uhlenbeck's result we have to assume that the reader is familiar 
  with the notion of {\em conjugate points\/} and with the following facts (see,
  e.g., Perlick \cite{Perlick2000}). The totality of all conjugate points, along any lightlike 
  geodesic issuing from a point $p$ into the past, makes up the {\em caustic\/} 
  of the past light-cone of $p$. A lightlike geodesic is said to have 
  (Morse) index $k$ if it has $k$ conjugate points in its interior; here and in the following every
  conjugate point has to be counted with its multiplicity. For a lightlike geodesic 
  with two end-points, the index is always finite. It is our goal to estimate 
  the number of past-pointing lightlike geodesics of index $k$ from a point $p$ 
  to a timelike curve $\gamma$ that does not meet the caustic of the past 
  light-cone of $p$. The latter condition is generically satisfied in the 
  sense that, for any $\gamma$, the set of all points $p$ for which it is 
  true is dense in $M$. This condition makes sure that the past-pointing 
  lightlike geodesics from $p$ to $\gamma$ are countable, i.e., it excludes 
  gravitational lensing situations where the observer sees a continuum of 
  images such as an Einstein ring.
  
  As another preparation, we recall how the \emph{Betti numbers} $B_k$ of the 
  \emph{loop space} $L(M)$ of a connected topological space $M$ are defined. 
  As a realization of $L(M)$ one may take the space of all continuous curves 
  between any two fixed points in $M$. The $k$th Betti number $B_k$ is 
  formally defined as the dimension of the $k$-th homology space of $L(M)$ 
  with coefficients in a field $\mathbb{F}$. (For our purpose we may choose 
  $\mathbb{F} = \mathbb{R}$.) 
  Roughly speaking, $B_0$ counts the connected components of $L(M)$
  and $B_k$, for $k>0$, counts those ``holes'' in $L(M)$ that prevent
  a $k-$sphere from being a boundary. If the reader is not familiar with 
  Betti numbers he or she may consult e.g. \cite{Frankel1997}.

  After these preparations Uhlenbeck's result that we want to use later in this 
  paper can now be phrased in the following way.

\begin{theorem}\label{theo:Uh}
  {\em (Uhlenbeck \cite{Uhlenbeck1975})}
  Consider a globally hyperbolic spacetime $(M,g)$ that admits an orthogonal 
  splitting $(\ref{eq:product}), (\ref{eq:globhyp})$ satisfying the metric growth 
  condition. Fix a point $p \in M$ and a smooth timelike curve $\gamma : {\mathbb{R}}
  \longrightarrow M$ which, in terms of the above-mentioned orthogonal splitting,
  takes the form $\gamma ( \tau ) = \big( \beta ( \tau ) , \tau \big)$, with a curve 
  $\beta : {\mathbb{R}} \longrightarrow S$. Moreover, assume that $\gamma$ does
  not meet the caustic of the past light-cone of $p$ and that for some sequence 
  $(\tau _i )_{i \in {\mathbb{N}}}$ with $\tau _i \rightarrow
  - \infty$ the sequence $\big( \beta (\tau _i ) \big) {}_{i \in {\mathbb{N}}}$ 
  converges in $S$. Then the Morse inequalities 
\begin{equation}\label{eq:Morseineq}
  N_k \ge B_k \;  \qquad {\text{for all}} \quad k \in {\mathbb{N}}_0
\end{equation}
  and the Morse relation
\begin{equation}\label{eq:Morserel}
  \sum_{k=0}^{\infty} (-1)^k N_k =
  \sum_{k=0}^{\infty} (-1)^k B_k 
\end{equation}
  hold true, where $N_k$ denotes the number of past-pointing lightlike geodesics with
  index $k$ from $p$ to $\gamma$, and $B_k$ denotes the $k$-th Betti number of the loop 
  space of $M$.
\end{theorem}
\begin{proof}
  See Uhlenbeck \cite{Uhlenbeck1975}, \S 4 and Proposition 5.2.
\end{proof}

  Please note that the convergence condition on $\big( \beta (\tau _i ) \big) 
  {}_{i \in {\mathbb{N}}}$ is certainly satisfied if $\beta$ is confined to a 
  compact subset of $S$, i.e., if $\gamma$ stays in a spatially compact set.

  The sum on the right-hand side of (\ref{eq:Morserel}) is, by definition, the
  Euler characteristic $\chi$ of the loop space of $M$. Hence, (\ref{eq:Morserel})
  can also be written in the form
\begin{equation}\label{eq:N+-}
  N_+ - N_- = \chi \, ,
\end{equation}
  where $N_+$ (respectively $N_-$) denotes the number of past-pointing lightlike
  geodesics with even (respectively odd) index from $p$ to $\gamma$.

  The Betti numbers of the loop space of $M=S \times {\mathbb{R}}$ are, of
  course, determined by the topology of $S$. Three cases are to be distinguished.

\noindent
  {\bf Case A}: $M$ is not simply connected. Then the loop space 
  of $M$ has infinitely many connected components, so $B_0 = \infty$. In this
  situation (\ref{eq:Morseineq}) says that $N_0 = \infty$, i.e., that there 
  are infinitely many past-pointing lightlike geodesics from $p$ to $\gamma$ 
  that are free of conjugate points.

\noindent
  {\bf Case B}: $M$ is simply connected but not contractible 
  to a point. Then for all but finitely many $k \in {\mathbb{N}}_0$ we have
  $B_k > 0$. This was proven in a classical paper by Serre \cite{Serre1951}, 
  cf. McKenzie \cite{McKenzie1985}. In this situation (\ref{eq:Morseineq}) implies 
  $N_k > 0$ for all but finitely many $k$. In other words, for almost 
  every positive integer $k$ we can find a past-pointing lightlike geodesic 
  from $p$ to $\gamma$ with $k$ conjugate points in its interior. Hence, 
  there must be infinitely many past-pointing lightlike geodesics from 
  $p$ to $\gamma$ and the caustic of the past light-cone of $p$ must be 
  complicated enough such that a past-pointing lightlike geodesic from $p$ 
  can intersect it arbitrarily often.  

\noindent
  {\bf Case C}: $M$ is contractible to a point. Then the loop space of 
  $M$ is contractible to a point, i.e., $B_0 =1$ and $B_k = 0$ for $k > 0$. 
  In this case (\ref{eq:N+-}) takes the form $N_+ - N_-
  = 1$ which implies that the total number $N_+ + N_- = 2 N_- + 1$
  of past-pointing lightlike geodesics from $p$ to $\gamma$ is (infinite
  or) odd.

  The domain of outer communication of a Kerr-Newman black hole has topology
  $S^2 \times {\mathbb{R}}^2$ which is simply connected but not contractible
  to a point. So it is Case B we are interested in when applying Uhlenbeck's
  result to the Kerr-Newman spacetime. 


\section{Centrifugal and Coriolis force in the Kerr-Newman 
spacetime}\label{sec:centrifugal}

  The Kerr-Newman metric is given in Boyer-Lindquist coordinates (see, e.g., 
  Misner, Thorne and Wheeler \cite{MisnerThorneWheeler1973}, p.877) by
\begin{equation}\label{eq:kerr}
  g =   - \frac{\Delta}{\rho ^2} \, \big( \, dt \, - \, 
  a \, \mathrm{sin} ^2 \vartheta \, d \varphi \big) ^2 \, + \,
  \frac{\mathrm{sin} ^2 \vartheta}{\rho ^2} \, \big(
  (r^2 + a^2) \, d \varphi \, - \, a \, dt \, \big) ^2 \, + \, 
  \frac{\rho ^2}{\Delta} \, dr^2 \,  + \, \rho ^2 \, d \vartheta ^2 \, ,
\end{equation}
  where $\rho$ and $\Delta$ are defined by
\begin{equation}\label{eq:rhodelta}
  \rho ^2 = r^2 + a^2 \, {\mathrm{cos}} ^2 \vartheta
  \quad \text{and} \quad 
  \Delta = r^2 - 2mr + a^2 + q^2 \, ,
\end{equation}
  and $m$, $q$ and $a$ are real constants.  We shall assume throughout that 
\begin{equation}\label{eq:ma}
  0 \, < \, m \: , \quad 0 \, \le \, a \: , \quad \sqrt{a^2 + q ^2} \, \le \, m \, .
\end{equation}
  In this case, the Kerr-Newman metric describes the spacetime around a rotating
  black hole with mass $m$, charge $q$, and specific angular momentum $a$. The 
  Kerr-Newman metric (\ref{eq:kerr}) contains the Kerr metric ($q=0$), the
  Reissner-Nordstr{\"om} metric ($a=0$) and the Schwarzschild metric ($q=0$ and $a=0$)
  as special cases which are all discussed, in great detail, in Chandrasekhar \cite{Chandrasekhar1983}; 
  for the Kerr metric we also refer to O'Neill \cite{ONeill1995}.  

  By (\ref{eq:ma}), the equation $\Delta = 0$ has two real roots,
\begin{equation}\label{eq:hor}
  r_{\pm} = m \pm \sqrt{ m^2 - a^2 - q ^2} \, ,
\end{equation}
  which determine the two horizons. We shall restrict to the region 
\begin{equation}\label{eq:M+}
  M_+ : \quad r_+ < r < \infty \, ,
\end{equation}
  which is usually called the {\em domain of outer communication\/} of the Kerr-Newman 
  black hole. On $M_+$, the coordinates $\varphi$ 
  and $\vartheta$ range over $S^2$, the coordinate $t$ ranges over ${\mathbb{R}}$, 
  and the coordinate $r$ ranges over an open interval which is diffeomorphic to 
  ${\mathbb{R}}$; hence $M_+ \simeq S^2 \times {\mathbb{R}}^2$.

  From now on we will consider the spacetime $(M_+,g)$, where 
  $g$ denotes the restriction of the Kerr-Newman metric (\ref{eq:kerr}) with (\ref{eq:ma}) 
  to the domain $M_+$ given by (\ref{eq:M+}). For the sake of brevity, we will 
  refer to $(M_+,g)$ as to the {\em exterior Kerr-Newman spacetime}. As a matter of 
  fact, $(M_+,g)$ is a globally hyperbolic spacetime; the Boyer-Lindquist time
  coordinate $t$ gives a foliation of $M_+$ into Cauchy surfaces $t = {\mathrm{constant}}$.
  Together with the lines perpendicular to these surfaces, we get an orthogonal
  splitting of the form (\ref{eq:globhyp}). Observers with worldlines perpendicular 
  to the surfaces $t = {\mathrm{constant}}$ are called {\em zero-angular-momentum
  observers\/} or \emph{locally non-rotating observers}. In contrast to the 
  worldlines perpendicular to the surfaces $t = {\mathrm{constant}}$, the integral 
  curves of the Killing vector field $\partial _t$ are {\em not\/} timelike on all 
  of $M_+ \,$; they become spacelike inside the socalled {\em ergosphere\/} which 
  is characterized by the inequality $\Delta < a^2 \mathrm{sin}^2 \vartheta$. 
  For $a \neq 0$ it is impossible to find a Killing vector field 
  which is timelike on all of $M_+$; in this sense, the exterior Kerr-Newman 
  spacetime is {\em not\/} a stationary spacetime.

  In the rest of this section we discuss the notions of centrifugal
  force and Coriolis force for observers on circular orbits around the 
  axis of rotational symmetry in the exterior Kerr-Newman spacetime $(M_+,g)$. For 
  background information on these notions we refer to the work of
  Marek Abramowicz and his collaborators \cite{AbramowiczCarterLasota1988, 
  Abramowicz1990, AbramowiczNurowskiWex1993} which was
  mentioned already in the introduction. For our discussion it will be convenient 
  to introduce on $M_+$ the orthonormal basis 
\begin{gather}
  E_0 = \frac{1}{\rho \, \sqrt{\Delta}} \Big( (r^2 + a^2) \partial _t 
  + a \partial _{\varphi} \Big) \: , 
\nonumber
\\
  E_1 = \frac{1}{\rho \, {\mathrm{sin}} \, \vartheta} \, 
  \big( \partial _{\varphi} + a \, {\mathrm{sin}} ^2 \vartheta \, \partial _t \big) \: , 
\label{eq:E}
\\
  E_2 = \frac{1}{\rho} \, \partial _{\vartheta} \: , 
\qquad
  E_3 = \frac{\sqrt{\Delta}}{\rho} \, \partial _r \: , 
\nonumber
\end{gather}
  whose dual basis is given by the covector fields
\begin{gather}
  -g(E_0 , \, \cdot \, ) = \frac{\sqrt{\Delta}}{\rho} \, \big( \, dt \, - \, 
  a \, {\mathrm{sin}}^2 \vartheta \, d \varphi \, \big) \: , 
\nonumber
\\
  g(E_1 , \, \cdot \, ) = \frac{\mathrm{sin} \, \vartheta}{\rho} \, 
  \big( \, (r^2+a^2) \; d \varphi \, - \, a \, dt \, \big)  \: ,
\label{eq:gE}
\\
  g(E_2 , \, \cdot \, ) = \rho \, d \vartheta \: ,
\qquad
  g(E_3 ,  \, \cdot \, ) = \frac{\rho}{\sqrt{\Delta}} \, dr \: . 
\nonumber
\end{gather}
  Henceforth we refer to the integral curves of the timelike basis field 
  $E_0$ as to the worldlines of the {\em standard observers\/} in $(M_+,g)\,$. 
  For later purpose we list all non-vanishing Lie brackets of the $E_i$. 
\begin{gather}
  [E_0,E_2] \, = \, - \, \frac{a^2}{\rho ^3} \, 
  \mathrm{cos} \, \vartheta \, \mathrm{sin} \, \vartheta \, E_0 \; ,
\nonumber
\\
  [E_0,E_3] \, = \, \Big( \, \frac{r-m}{\rho \sqrt{\Delta}} \, - \, 
  \frac{r \sqrt{\Delta}}{\rho ^3} \, \Big) \, E_0 \, + \, 
  \frac{2 \, r \, a \, \mathrm{sin} \, \vartheta}{\rho ^3} \, E_1 \; ,
\nonumber
\\
  [E_1,E_2] \, = \, 
  \frac{( \rho ^2 + a^2 \mathrm{sin} ^2 \vartheta ) \, 
  \mathrm{cos} \, \vartheta}{\rho ^3 \, \mathrm{sin} \, \vartheta} \, E_1 \, - \,
  \frac{2 \, a \, \sqrt{\Delta} \, \mathrm{cos} \, \vartheta}{\rho ^3} \, E_0 \; ,
\label{eq:Lie}
\\
  [E_1,E_3] \, = \, \frac{r \, \sqrt{\Delta}}{\rho ^3} \, E_1 \; ,
\nonumber
\\
  [E_2,E_3] \, = \, \frac{r \, \sqrt{\Delta}}{\rho ^3} \, E_2 \, + \,
  \frac{a^2 \mathrm{cos} \, \vartheta \, \mathrm{sin} \, \vartheta}{\rho ^3} 
  \, E_3 \; . 
\nonumber
\end{gather} 

  For every $v \in [0,1\, [ \, $, the integral curves of the vector field
\begin{equation}\label{eq:U}
  U = \frac{E_0 \pm v \, E_1}{\sqrt{1-v^2}}
\end{equation}
  can be interpreted as the worldlines of observers who circle along the $\varphi$-lines
  around the axis of rotational symmetry of the Kerr-Newman spacetime.  The number $v$ 
  gives the velocity (in units of the velocity of light) of these observers with respect 
  to the standard observers. For the upper sign in (\ref{eq:U}), the motion relative 
  to the standard observers is in the positive $\varphi$-direction and thus co-rotating 
  with the black hole (because of our assumption $a \ge 0$), for the negative sign it 
  is in the negative $\varphi$-direction and thus counter-rotating. Please 
  note that $g(U,U)=-1$, which demonstrates that the integral curves of $U$ are 
  parametrized by proper time.  

  In general, $U$ is non-geodesic, $\nabla _U U \neq 0$, i.e., one needs a thrust to 
  stay on an integral curve of $U$. Correspondingly, relative to a $U$-observer a 
  freely falling particle undergoes an ``inertial acceleration'' measured by 
  $- \nabla _U U$. To calculate this quantity, we write 
\begin{equation}\label{eq:compacc}
  -g(\nabla _U U,E_i) = 
  -U g(U,E_i) + g(U ,\nabla _U E_i) = -U g(U,E_i)+ g(U,[U,E_i]) \, . 
\end{equation}
  The first term on the right-hand side vanishes, and the second term can be
  easily calculated with the help of (\ref{eq:U}) and (\ref{eq:Lie}), 
  for i=0,1,2,3. We find
\begin{equation}\label{eq:delUU}
  -g(\nabla _U U , \, \cdot \, ) = 
  \, A_{\mathrm{grav}} \, + \, A_{\mathrm{Cor}} \, + \, A_{\mathrm{cent}}
\end{equation}  
  where the covector fields 
\begin{gather}
\label{eq:Agrav}
  A_{\mathrm{grav}} = \frac{\Delta \, r - \rho ^2 (r-m)}{\rho^2 \Delta} \, dr
  + \frac{a^2}{\rho^2} \; {\mathrm{sin}} \, \vartheta \; 
  {\mathrm{cos}}\, \vartheta \; d \vartheta \; ,
\\
\label{eq:ACor}
  A_{\mathrm{Cor}}= \, \pm \, \frac{v}{(1-v^2)} \; \frac{2 \, a \, \sqrt{\Delta}}{\rho ^2} \;
  \Big( \, \frac{r}{\Delta} \, {\mathrm{sin}} \, \vartheta \; dr \; + 
  \; {\mathrm{cos}}\, \vartheta \; d \vartheta \, \Big) \; , 
\\
\label{eq:Acent}
  A_{\mathrm{cent}}= \frac{v^2}{(1-v^2)} \;  
  \Big( \, \frac{2 \, r \, \Delta-\rho^2 (r-m)}{\rho^2 \Delta} \,  dr + \frac{( \, 
  \rho^2+ 2 \, a^2 {\mathrm{sin}}^2 \vartheta \, ){\mathrm{cos}}\, \vartheta}{
  \rho ^2 \, {\mathrm{sin}}\, \vartheta}
  \; d\vartheta \, \Big)  
\end{gather}
  give, respectively, the gravitational, the Coriolis, and the centrifugal acceleration
  of a freely falling particle relative to the $U$-observers. (Multiplication with the 
  particle's mass gives the corresponding ``inertial force''.) Here the decomposition of 
  the total inertial acceleration into its three contributions is made according to
  the same rule as in Newtonian mechanics: The gravitational acceleration is independent
  of $v$, the Coriolis acceleration is odd with respect to $v$, and the centrifugal acceleration
  is even with respect to $v$. In \cite{FoertschHassePerlick2003} it was shown that, according to this rule,
  gravitational, Coriolis and centrifugal acceleration are unambiguous whenever a timelike
  2-surface with a timelike vector field has been specified; here we apply this procedure
  to each 2-surface $(r, \vartheta ) = \mathrm{constant}$ with the timelike vector field 
  $E_0$.

  Up to the positive factor $v  / (1-v^2)$, the sum of Coriolis and centrifugal 
  acceleration is equal to 
\begin{equation}\label{eq:Z}
\begin{split}
  Z_{\pm}(v) =  \pm \, \frac{2 \, a \, \sqrt{\Delta}}{\rho ^2} \;
  & \Big( \, \frac{r}{\Delta} \, {\mathrm{sin}} \, \vartheta \; dr \; + 
  \; {\mathrm{cos}}\, \vartheta \; d \vartheta \, \Big)
\\
  + \, v \, 
  \Big( \, \frac{2 \, r \, \Delta-\rho^2 (r-m)}{\rho^2 \, \Delta} \,  d &r + 
  \frac{(\, \rho^2+ 2 \, a^2 \, {\mathrm{sin}}^2 \vartheta \, ) \,
  {\mathrm{cos}}\, \vartheta}{\rho ^2 \, {\mathrm{sin}}\, \vartheta}
  \; d\vartheta \, \Big) \; .
\end{split}
\end{equation}
  If we exclude the Reissner-Nordstr{\"o}m case $a=0$, the Coriolis force 
  dominates the centrifugal force for small $v$. To investigate the behavior 
  for $v$ close to the velocity of light, we consider the limit $v \rightarrow 1$. 
  By a straight-forward calculation we find that
\begin{equation}\label{eq:ZPsi}
  Z_{\pm} (v) \: {\underset{v \to 1}{\longrightarrow}}  \: 
  \frac{\mathrm{sin} \, \vartheta}{\rho ^2 \sqrt{\Delta}} \, 
  \big( r^2 + a^2 \pm a \, \sqrt{\Delta} \; {\mathrm{sin}}\, \vartheta \, \big)^2
  \; d \Psi_{\pm} \; ,
\end{equation}
  where 
\begin{equation}\label{eq:dPsi}
\begin{split}
  d \Psi _{\pm} \; & =  \;
  \frac{
  2\, r \, \Delta -(r-m) \, \rho^2  
  \pm 2 \, a \, r \, \sqrt{\Delta} \; {\mathrm{sin}}\, \vartheta  
  }{
  \sqrt{\Delta} \; {\mathrm{sin}} \, \vartheta \, 
  \big( r^2 +a^2 \pm \, a \, \sqrt{\Delta} \; {\mathrm{sin}} \, \vartheta \, \big)^2
  } \; \; dr
\\
  & + \; 
  \frac{
  \big( \rho^2 + 2 \, a^2 {\mathrm{sin}}^2 \vartheta \pm 
  2 \, a \, \sqrt{\Delta} \; {\mathrm{sin}} \, \vartheta \, \big) \, \sqrt{\Delta} \;
  {\mathrm{cos}}\, \vartheta
  }{
  {\mathrm{sin}}^2 \vartheta \, 
  \big( r^2 +a^2 \pm \, a \sqrt{\Delta} \; {\mathrm{sin}} \, \vartheta \, \big)^2
  } \; \; d \vartheta
\end{split}
\end{equation}
  is the differential of the function
\begin{equation}\label{eq:Psi}
  \Psi_{\pm} 
  = \frac{
  - \frac{1}{{\mathrm{sin}} \, \vartheta} \mp \frac{a}{\sqrt{\Delta}}
  }{
  \frac{r^2+a^2}{\sqrt{\Delta}} \pm a \, {\mathrm{sin}} \, \vartheta
  } \: .
\end{equation}
  Because of $\, {\mathrm{sin}} \, \vartheta \,$ in the denominator, both
  $\Psi_-$ and $\Psi_+$ are singular along the axis. $\Psi_+$ is negative 
  on all of $M_+$ whereas $\Psi_-$ is negative outside and positive inside 
  the ergosphere. 

\begin{figure}
\vspace{-0.5cm}
\centerline{\epsfig{figure=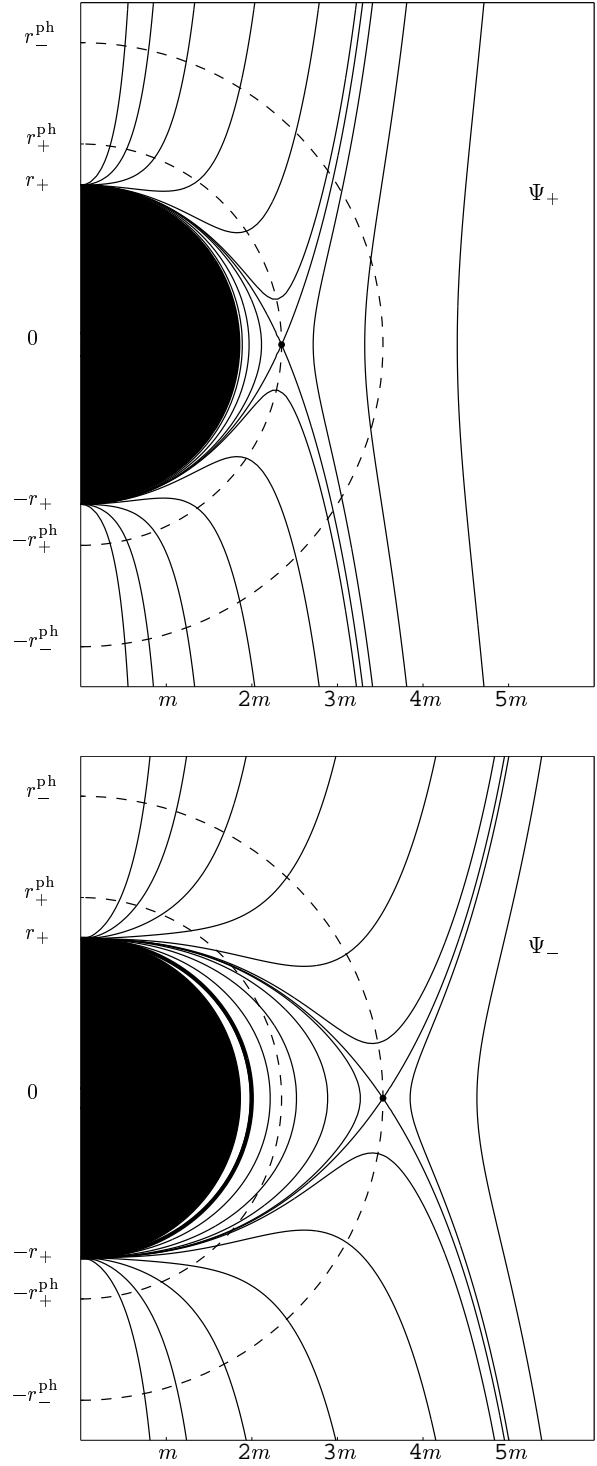, width=16cm}}
\vspace{-2.1cm}
\caption{ The surfaces $\Psi _+ = {\mathrm{constant}}$ (top) and
  $\Psi_- = {\mathrm{constant}}$ (bottom) are drawn here for the case 
  $q=0$ and $a = 0.5 \, m \,$. The picture shows the (half-)plane $(\varphi ,t)
  = \mathrm{constant}$, with $r \, \mathrm{sin} \, \vartheta$ on the horizontal
  and $r \, \mathrm{cos} \, \vartheta$ on the vertical axis. The spheres of 
  radius $r_+^{\mathrm{ph}}$ and $r_-^{\mathrm{ph}}$ are indicated by dashed 
  lines; they meet the equatorial plane in the photon circles. The boundary of 
  the ergosphere coincides with the surface $\Psi_-=0$ and is indicated in
  the bottom figure by a thick line; it meets the equatorial plane at 
  $r = 2m$. \label{fig:Psi}}
\end{figure}

  From (\ref{eq:ZPsi}) we read that, in the limit $v \rightarrow 1$, the sum of 
  Coriolis and centrifugal force is perpendicular to the surfaces 
  $\Psi _{\pm} = {\mathrm{constant}}$ and points in the direction of increasing
  $\Psi_{\pm}$. In this limit, we may thus view the function $\Psi_+$ (or
  $\Psi_-$, resp.) as a Coriolis-plus-centrifugal potential for co-rotating
  (or counter-rotating, resp.) observers. The surfaces $\Psi _{\pm} = 
  {\mathrm{constant}}$ are shown in Figure \ref{fig:Psi}.

  It is not difficult to see that $\Psi_{\pm}$ is independent of the 
  family of observers with respect to which the inertial accelerations
  have been defined, as long as their 4-velocity is a linear combination of 
  $\partial _t$ and $\partial _{\varphi}$. We have chosen the standard 
  observers; a different choice would lead to different formulas for the 
  inertial accelerations (\ref{eq:Agrav}), (\ref{eq:ACor}) and (\ref{eq:Acent}), 
  but to the same $\Psi_{\pm}$. For the sake of comparison, the reader may
  consult Nayak and Vishveshwara \cite{NayakVishveshwara1996} where the
  inertial accelerations are calculated with respect to the zero angular momentum
  observers. Also, it should be mentioned that the potentials 
  $\Psi_+$ and $\Psi_-$, or closely related functions, have been used already 
  by other authors. The quantities $\Omega _{c \pm}$, e.g., introduced by de 
  Felice and Usseglio-Tomasset \cite{deFeliceUsseglio1991} in  their analysis 
  of physical effects related to centrifugal force reversal in the equatorial 
  plane of the Kerr metric, are related to our potentials by $\Omega _{c \pm} = 
  \mp \Psi_{\pm} |_{\vartheta = \pi /2}$.

  In the Reissner-Nordstr{\"o}m case $a=0$, the Coriolis acceleration (\ref{eq:ACor}) 
  vanishes identically and
\begin{equation}\label{eq:Psi0}
  \Psi = \Psi_+ = \Psi_- = 
  - \frac{\sqrt{\, r^2 \, - \, 2 \, m \, r \, + \, q^2 \,}}{
  r^2 \, {\mathrm{sin}} \, \vartheta} 
\end{equation}
  is a potential for the centrifugal acceleration in the sense that $A_{\mathrm{cent}}$ 
  is a multiple of $d \Psi$. In this case, the surfaces $\Psi = {\mathrm{constant}}$
  coincide with what Abramowicz \cite{Abramowicz1990} calls the \emph{von Zeipel cylinders}. 
  Abramowicz's Figure 1 in \cite{Abramowicz1990}, which shows the von Zeipel 
  cylinders in the Schwarzschild spacetime, coincides with the $a \to 0$ limit of our 
  Figure \ref{fig:Psi}, which shows the surfaces $\Psi_+ = \mathrm{constant}$ and 
  $\Psi_- = \mathrm{constant}$ in the Kerr spacetime. (The notion of von Zeipel cylinders 
  has also been defined in the Kerr metric, see \cite{KozlowskiJaroszynskiAbramowicz1978}, 
  for observers of a specified angular velocity. However, this angular-velocity-dependent 
  von Zeipel cylinders are not related to the potentials $\Psi_+$ and $\Psi_-$ in the Kerr 
  spacetime.)
  
  By construction, the function $\Psi _{\pm}$ has the following property. If
  we send a lightlike geodesic tangential to a $\varphi$-line in the positive
  (respectively negative) $\varphi$-direction, it will move away from this
  $\varphi$-line in the direction of the negative gradient of $\Psi _+$ (respectively
  $\Psi _ -$). Thus, each zero of the differential $d \Psi _+$ (respectively
  $d \Psi _-$) indicates a co-rotating (respectively counter-rotating) circular
  lightlike geodesic, i.e., a ``photon circle''. By (\ref{eq:dPsi}), 
  $d \Psi _{\pm}$ vanishes if 
\begin{equation}\label{eq:crit}
  {\mathrm{cos}}\, \vartheta \, =0 \quad \text{and} \quad
  2 \, r \, \Delta - (r-m) \, \rho^2
  \pm 2 \, a \, r \, \sqrt{\Delta} \; {\mathrm{sin}} \, \vartheta \, =0 \; .
\end{equation}
  By writing $\Delta$ and $\rho^2$ explicitly, we see that (\ref{eq:crit}) is true
  at $\vartheta = \pi /2$ and $r = r_{\pm}^{\mathrm{ph}}$, where 
  $r_{\pm}^{\mathrm{ph}}$ is defined by the equation
\begin{equation}\label{eq:rph}
  \big( r_{\pm}^{\mathrm{ph}} \big)^2  - 3 \, m \, r_{\pm}^{\mathrm{ph}}
  + 2 \, a^2 \, + \, 2 \, q ^2 \, = \, \mp \; 
  2 \, a \, \sqrt{ \big( r_{\pm}^{\mathrm{ph}} \big)^2  - 
  2 \, m \,  r_{\pm}^{\mathrm{ph}} \, + \, a^2 \, + \, q ^2 \,} \, .
\end{equation}
  For $0 \, < \, \sqrt{a^2+q^2} \, < \, m \,$, (\ref{eq:rph}) has exactly one 
  solution for each sign which satisfies
\begin{equation}\label{eq:order}
  r_+ \, < \, r_+^{\mathrm{ph}} \, < \, 
  \frac{3 \, m}{2} \, + \, \sqrt{ \frac{9 m ^2}{4} \, - \, 2 \, q ^2 \, } \, < \,
  r_-^{\mathrm{ph}}  \, < \, \, 2 \, m \, + \, 2 \, \sqrt{m^2 -q^2} \, .
\end{equation}
  So there is exactly one co-rotating 
  photon circle in $M_+$, corresponding to the critical point of 
  $\Psi_+$ at $r_+^{\mathrm{ph}}$, and exactly one counter-rotating photon 
  circle in $M_+$, corresponding to the critical point of $\Psi_-$ 
  at $r_-^{\mathrm{ph}}$, see Figure \ref{fig:Psi}. (The relation of photon
  circles to centrifugal-plus-Coriolis force in the limit $v \to 1$ is also
  discussed by Stuchlik, Hledik and Jur{\'a}n \cite{StuchlikHledikJuran2000};
  note, however, that their work is restricted to the equatorial plane of
  the Kerr-Newman spacetime throughout.)    
  In the Reissner-Nordstr{\"o}m case, $a=0$, we have $r_+^{\mathrm{ph}} = 
  r_-^{\mathrm{ph}} = \frac{3m}{2} +  \sqrt{\frac{9 m^2}{4} - 2 q^2}$ 
  (cf., e.g., Chandrasekhar \cite{Chandrasekhar1983}, p.218). If we keep $m$ and $q$ fixed 
  and vary $a$ from $0$ to the extreme value $\sqrt{m^2-q^2}$, $r_+^{\mathrm{ph}}$ 
  decreases from $\frac{3}{2}m +  \sqrt{\frac{9 \, m^2}{4} -2 q^2}$
  to $m$ whereas $r_-^{\mathrm{ph}}$ increases from $\frac{3}{2}m +  
  \sqrt{\frac{9 \, m^2}{4} - 2 q^2}$ to $2m + 2 \sqrt{m^2 - q^2}$. As an aside, we 
  mention that, although $r_+^{\mathrm{ph}}$ and $r_+$ both go to $m$ in the
  extreme case, the proper distance between the co-rotating photon circle at
  $r_+^{\mathrm{ph}}$ and the horizon at $r_+$ does not go to zero; for the 
  case $q=0$ this surprising feature is discussed in Chandrasekhar \cite{Chandrasekhar1983},
  p. 340. 


  From (\ref{eq:dPsi}) we can read the sign of $\partial _r \Psi _{\pm}$ at 
  each point. We immediately find the following result.

\begin{figure}[t]
  \centerline{\epsfig{figure=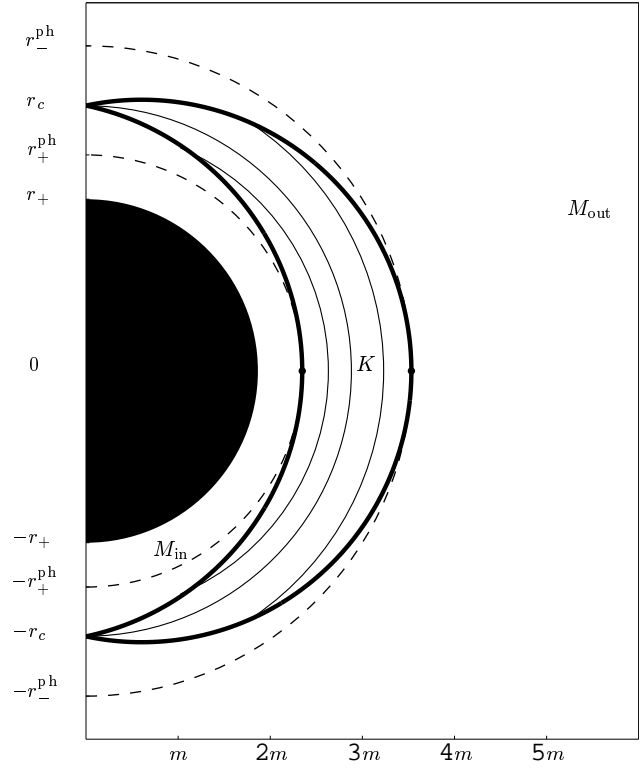, width=15.5cm}}
   \vspace{-5.5cm}
  \caption{ The regions $M_{\mathrm{in}}$, $K$ and $M_{\mathrm{out}}$ defined
    in Proposition \ref{prop:centri} are shown here for the case $q=0$ and 
    $a = 0.5 \, m \,$. Again, as in Figure \protect\ref{fig:Psi}, we plot $r \, \mathrm{sin} 
    \, \vartheta$ on the horizontal and $r \, \mathrm{cos} \, \vartheta$ on the 
    vertical axis. Some of the spherical lightlike geodesics that fill the photon 
    region $K$ are indicated. $K$ meets the equatorial plane in the photon circles 
    at $r = r_+^{\mathrm{ph}}$ and $r=r_-^{\mathrm{ph}}$ and the axis at radius $r_c$
    given by $r_c^3 \, - \, 3 \, r_c^2 \, m \, + \, r_c \, ( \, a^2  \, + 2 \, q^2 \, )
    \, + \, a^2 \, m \, = \, 0 \, $. -- This picture can also be found 
    as Figure 21 in the online article \protect\cite{Perlick2004}.}\label{fig:K} 
\end{figure}

\begin{proposition}\label{prop:centri}
  Decompose the exterior Kerr spacetime into the sets 
\begin{eqnarray}
  M_{\mathrm{in}} \; : \qquad & 
  2 \, r \, \Delta - (r-m) \, \rho ^2 \: <
  \: - \, 2 \, a \, r \, \sqrt{\Delta} \,  {\mathrm{sin}} \, \vartheta 
\label{eq:Min}
\\
  K \; \; : \qquad & 
  \, - \, 2 \, a \, r \, \sqrt{\Delta} \,  {\mathrm{sin}} \, \vartheta  \: \le
  \: 2 \, r \, \Delta - (r-m) \, \rho ^2 \: \le
  \: 2 \, a \, r \, \sqrt{\Delta} \,  {\mathrm{sin}} \, \vartheta 
\label{eq:K}
\\
  M_{\mathrm{out}} \; : \qquad & 
  2 \, a \, r \, \sqrt{\Delta} \,  {\mathrm{sin}} \, \vartheta  \: <
  \: 2 \, r \, \Delta - (r-m) \, \rho ^2 \, ,
\label{eq:Mout}
\end{eqnarray}
  so $M_+ = M_{\mathrm{in}} \cup K \cup M_{\mathrm{out}}$, see 
  Figure $\ref{fig:K}$. Then 
\begin{eqnarray}
  \partial _r \Psi _+ < 0 \quad \text{and} \quad \partial _r \Psi _- < 0
  \quad & \text{on } \, M_{\mathrm{in}}  \, ,
\label{eq:in}
\\
  \partial _r \Psi _+ < 0 \quad \text{and} \quad \partial _r \Psi _- > 0
  \quad  & \text{on the interior of } \, K  \, ,
\label{eq:trans}
\\
  \partial _r \Psi _+ > 0 \quad \text{and} \quad \partial _r \Psi _- > 0
  \quad & \text{on } \, M_{\mathrm{out}}  \, .
\label{eq:out}
\end{eqnarray}
\end{proposition}

  The inequality $\partial _r \Psi _{\pm} > 0$ is true for both signs 
  if and only if, for $v$ sufficiently large, the sum of Coriolis and 
  centrifugal force is pointing in the direction of increasing $r$ for 
  co-rotating and counter-rotating observers. An equivalent condition is that 
  the centrifugal force points in the direction of increasing $r$ and dominates 
  the Coriolis force for $v$ sufficiently large. This is the situation we are familiar 
  with from Newtonian physics. According to Proposition \ref{prop:centri}, however, in 
  the Kerr-Newman spacetime this is true only in the region $M_{\mathrm{out}}$.
  In the interior of the intermediate region $K$ the direction of 
  centrifugal-plus-Coriolis force for large $v$ is reversed for 
  counter-rotating observers while still normal for co-rotating observers.  
  In the region $M_{\mathrm{in}}$, finally, it is reversed both for
  co-rotating and for counter-rotating observers.

  The relevance of the sets $M_{\mathrm{out}}$, $M_{\mathrm{in}}$ and $K$
  in view of lightlike geodesics is demonstrated in the following proposition.

\begin{proposition}\label{prop:convex}
\begin{itemize}
\item[\emph{(a)}]
  In the region $M_{\mathrm{out}}$, the radius coordinate $r$ cannot have
  other extrema than strict local minima along a lightlike geodesic.
\item[\emph{(b)}]
  In the region $M_{\mathrm{in}}$, the radius coordinate $r$ cannot have
  other extrema than strict local maxima along a lightlike geodesic.
\item[\emph{(c)}]
  Through each point of $K$ there is a spherical lightlike 
  geodesic. $($Here ``spherical'' means that the geodesic is completely
  contained in a sphere $r = \mathrm{constant}.)$  
\end{itemize}
\end{proposition}
\begin{proof}
  Let $X$ be a lightlike and geodesic vector field on $(M_+,g)$, i.e., $g(X,X)=0$ and 
  $\nabla_X X = 0$. To prove (a) and (b), we have to demonstrate that the implication
\begin{equation}\label{eq:Xpos}
  X r = 0 \quad \Rightarrow \quad XXr > 0
\end{equation}
  is true at all points of $M_{\mathrm{out}}$ and that the implication
\begin{equation}\label{eq:Xneg}
  X r = 0 \quad \Rightarrow \quad XXr < 0
\end{equation}
  is true at all points of $M_{\mathrm{in}}$.  Here $X r$ is 
  to be read as ``the derivative operator $X$ applied to the function $r$''.
  The condition $\nabla_X X=0$ implies
\begin{equation}\label{eq:XXr}
  XXr = X dr(X)= X \, \Big( \, \frac{\sqrt{\Delta}}{\rho} \, g (E_3, X ) \, \Big) 
  \, = \,  \frac{\sqrt{\Delta}}{\rho} \, g \big( \nabla _X E_3 , X \big) +
  \Big( \, X \, \frac{\sqrt{\Delta}}{\rho} \, \Big) \, g(E_3,X) \; ,
\end{equation}
  where we have used the basis vector field $E_3$ from (\ref{eq:E}) and (\ref{eq:gE}).
  Using these orthonormal basis vector fields, we can write $X$ in the form
\begin{equation}\label{eq:XE}
  X \, = \, E_0 \, + \, {\mathrm{cos}} \, \alpha \; E_1 \, 
  + \, {\mathrm{sin}} \, \alpha \; E_2 
\end{equation}
  at all points where $Xr=0$. (A non-zero factor of $X$ is irrelevant because $X$
  enters quadratically into the right-hand side of (\ref{eq:XXr}).)
  Then (\ref{eq:XXr}) takes the form
\begin{gather}
  \frac{\rho}{\sqrt{\Delta}} \, XXr \, = \, 
  g \big( \nabla _{E_0} E_3 , E_0 \big) \, + \,  
  \mathrm{sin} \, \alpha \, \Big( \,
  g \big( \nabla _{E_2} E_3 , E_0 \big) \, + \, 
  g \big( \nabla _{E_0} E_3 , E_2 \big) \, \Big) \, + \,
\nonumber
\\
  \mathrm{cos} \, \alpha \, \Big( \,
  g \big( \nabla _{E_1} E_3 , E_0 \big) \, + \, 
  g \big( \nabla _{E_0} E_3 , E_1 \big) \, \Big) \, + \,
  \mathrm{sin}^2  \alpha \, 
  g \big( \nabla _{E_2} E_3 , E_2 \big) \, + \, 
\nonumber
\\
  \mathrm{cos} ^2 \alpha \, 
  g \big( \nabla _{E_1} E_3 , E_1 \big) \, = \, 
  g \big( [E_0 , E_3 ], E_0 \big) \, + \,  
\label{eq:XXrE}
\\
  \mathrm{sin} \, \alpha \, \Big( \,
  g \big( [ E_2 , E_3 ] , E_0 \big) \, + \, 
  g \big( [ E_0 , E_3 ] , E_2 \big) \, \Big) \, + \,
  \mathrm{cos} \, \alpha \, \Big( \,
  g \big( [ E_1 , E_3 ] , E_0 \big) \, + \, 
  g \big( [ E_0 , E_3 ] , E_1 \big) \, \Big) \, + \,
\nonumber
\\
  \mathrm{sin}^2  \alpha \, 
  g \big( [ E_2 , E_3 ] , E_2 \big) \, + \, 
  \mathrm{cos} ^2 \alpha \, 
  g \big( [ E_1 , E_3 ] , E_1 \big) \; . 
\nonumber
\end{gather}
  If we insert the Lie brackets from (\ref{eq:Lie}) we find
\begin{equation}\label{eq:alpha}
  \rho ^4 \, X X r \, = \, 
  2 \, r \, \Delta - (r-m) \, \rho^2 + 2 \, a \, r \, \sqrt{\Delta} \;
  {\mathrm{sin}} \, \vartheta \; {\mathrm{cos}} \, \alpha \, . 
\end{equation}
  Now we compare this expression with (\ref{eq:Min}), (\ref{eq:K})
  and (\ref{eq:Mout}).  
  If ${\mathrm{cos}} \, \alpha$ runs through all possible values from $-1$ to 1,
  the right-hand side of (\ref{eq:alpha}) stays positive on $M_{\mathrm{out}}$
  and negative on $M_{\mathrm{in}}$. This proves part (a) and part (b). 
  At each point of $K$ there is exactly one value of ${\mathrm{cos}} \, 
  \alpha$ such that the right-hand side of (\ref{eq:alpha}) vanishes. This 
  assigns to each point of $K$ a lightlike direction such that the integral 
  curves of the resulting direction field are spherical lightlike geodesics. 
  This proves part (c).
\end{proof} 

  In view of part (c) of Proposition \ref{prop:convex} we refer to the closed
  region $K$ as to the \emph{photon region} of the exterior Kerr-Newman 
  spacetime. Along each spherical lightlike geodesic in $K$ the 
  $\vartheta$-coordinate oscillates between extremal values $\vartheta _0$
  and $- \vartheta _0$, correponding to boundary points of $K$, see Figure
  \ref{fig:K}; the $\varphi$-coordinate either increases or decreases monotonically.
  In the Reissner-Nordstr{\"o}m case $a=0$, where (\ref{eq:Psi0}) is a 
  potential for the centrifugal force, the photon region $K$ shrinks 
  to the \emph{photon sphere} $r \, = \,  
  \frac{3}{2} m + \sqrt{ \frac{9 \, m^2}{4}  \, - \, 2 \, q ^2 \, }$
  and Proposition \ref{prop:centri} reduces to the known fact that centrifugal
  force reversal takes place at the photon sphere. 
  
  We end this section with a word of caution as to terminology. In part (c) of 
  Proposition \ref{prop:convex} we have refered to the set $r=\mathrm{constant}$ as 
  to a 'sphere'. This is indeed justified in the sense that, for each fixed $t$, 
  fixing the radius coordinate $r$ gives a two-dimensional submanifold of $M_+$ that 
  is diffeomorphic to the 2-sphere. Moreover, in our Figures \ref{fig:Psi} and 
  \ref{fig:K} the sets $r=\mathrm{constant}$ are represented as (meridional 
  cross-sections of) spheres. Note, however, that the Kerr-Newman metric does 
  \emph{not} induce an isotropic metric on these spheres (unless $a=0$), so they 
  are not 'round spheres' in the metrical sense.
   


\section{Multiple imaging in the Kerr-Newman spacetime}\label{sec:imaging}
  
  It is now our goal to discuss multiple imaging in the exterior Kerr-Newman spacetime 
  $(M_+,g)$. To that end we fix a point $p$ and a timelike curve $\gamma$ in $M_+$
  and we want to get some information about the past-pointing lightlike geodesics
  from $p$ to $\gamma$. The following proposition is an immediate consequence of 
  Proposition \ref{prop:convex}.

\begin{proposition}\label{prop:shell}
  Let $p$ be a point and $\gamma$ a timelike curve in the exterior Kerr-Newman 
  spacetime. Let
\begin{equation}\label{eq:shell}
  \Lambda \, : \quad r_a < r < r_b
\end{equation}
  denote the smallest spherical shell, with $r_+ \le r_a < r_b \le \infty$, 
  such that $p$, $\gamma$ and the region $K$ defined by {\em (\ref{eq:K})} 
  are completely contained in ${\overline{\Lambda}}\, ( \, = \, $closure of 
  $\Lambda$ in $M_+)$. Then all lightlike geodesics that join $p$ and 
  $\gamma$ are confined within ${\overline{\Lambda}}$.
\end{proposition}
\begin{proof}
  Along a lightlike geodesic that leaves and re-enters $\overline{\Lambda}$ the 
  radius coordinate $r$ must have either a maximum in the region $M_{\mathrm{out}}$ 
  or a minimum in the region $M_{\mathrm{in}}$. Proposition \ref{prop:convex} makes sure 
  that this cannot happen.
\end{proof}
  
  By comparison with Proposition \ref{prop:centri} we see that, among all spherical 
  shells whose closures in $M_+$ contain $p$ and $\gamma$, the shell $\Lambda$ of 
  Proposition \ref{prop:shell} is the smallest shell such that at all points of 
  the boundary of $\Lambda$ in $M_+$ the gradient of $\Psi _+$ and 
  the gradient of $\Psi_-$ are pointing in the direction away from $\Lambda$. 
  Based on Proposition \ref{prop:shell}, we will later see that there is a close
  relation between multiple imaging and centrifugal-plus-Coriolis force reversal 
  in the Kerr-Newman spacetime. 

  Proposition \ref{prop:shell} tells us to what region the lighlike geodesics between
  $p$ and $\gamma$ are confined, but it does not tell us anything about the number of
  these geodesics. To answer the latter question, we now apply Theorem \ref{theo:Uh}
  to the exterior Kerr-Newman spacetime $(M_+,g)$. 

\begin{proposition}\label{prop:infinite}
  Consider, in the exterior Kerr-Newman spacetime $(M_+,g)$, a point $p$ and a smooth 
  future-pointing timelike curve $\gamma : \; ] - \infty , \, \tau _a \; [ \; 
  \longrightarrow M_+$, with $- \infty < \tau _a \le \infty$, which is parametrized by 
  the Boyer-Lindquist time coordinate $t$, i.e., the $t$-coordinate of the point 
  $\gamma (\tau)$ is equal to $\tau$. Assume {\em (i)\/} that $\gamma$ 
  does not meet the caustic of the past light-cone of $p$, and {\em (ii)\/} that
  for $\tau \rightarrow - \infty$ the radius coordinate $r$ of the point $\gamma (\tau)$ 
  remains bounded and bounded away from $r_+$. $($The last condition means that 
  $\gamma (\tau)$ goes neither to infinity nor to the horizon for 
  $\tau \rightarrow - \infty$.$)$ Then there is an infinite sequence
  $(\lambda _n) _{n \in {\mathbb{N}}}$ of mutually different past-pointing 
  lightlike geodesics from $p$ to $\gamma$. For $n \rightarrow \infty$, the 
  index of $\lambda _n$ goes to infinity. Moreover, if we denote the point where
  $\lambda _n$ meets the curve $\gamma$ by $\gamma ( \tau _n )$, then $\tau _n
  \rightarrow - \infty$ for $n \rightarrow \infty$.
\end{proposition}
\begin{proof}
  We want to apply Theorem \ref{theo:Uh} to the exterior Kerr-Newman spacetime $(M_+,g)$.
  To that end, the first thing we have to find is an orthogonal splitting of the
  exterior Kerr-Newman spacetime that satisfies the metric growth condition. As in the 
  original Boyer-Lindquist coordinates the $t$-lines are not orthogonal to the
  surfaces $t = {\mathrm{constant}}$, we change to new coordinates
\begin{equation}\label{eq:coord}
  x^1 = r \, , \quad
  x^2 = \vartheta \, , \quad
  x^3 =  \varphi - u(r, \vartheta ) \, t \, , \quad
  t = t  \, , 
\end{equation}
with
\begin{equation}\label{eq:u}
  u(r,\vartheta ) \, = \, 
  \frac{2 \, m \, a \, r}{\rho ^2 \Delta + 2 \, m \, r \, (r^2+a^2)} \; .
\end{equation}
  Then the Kerr metric (\ref{eq:kerr}) takes the orthogonal splitting form 
  (\ref{eq:globhyp}), with  
\begin{equation}\label{eq:gij}
\begin{split}
  g_{ij}(x, t ) \, dx^i dx^j = 
  \rho ^2 \, \Big( \, \frac{dr^2}{\Delta} + d \vartheta ^2 \, \Big) +
  \qquad \qquad \qquad \qquad
\\
  \frac{{\mathrm{sin}}^2 \vartheta}{\rho^2} \, \big( \, (r^2 +a^2)^2 -
  \Delta \, a ^2 \, \mathrm{sin}^2 \vartheta \, \big) \, 
  \Big( \, t \, \big( \, \frac{\partial u (r, \vartheta)}{\partial r} \, dr + 
  \frac{\partial u (r, \vartheta)}{\partial \vartheta} \, d \vartheta \, \big) + 
  d x^3 \, \Big)^2
\end{split}
\end{equation}
and
\begin{equation}\label{eq:f}
  f(x, t ) = \frac{\rho ^2 \, \Delta}{
  (r^2+a^2)^2 - \Delta \, a ^2 \, \mathrm{sin} ^2 \vartheta} \; .
\end{equation}
  Clearly, if we restrict the range of the coordinates $x = (x^1,x^2,x^3)$ to a 
  compact set, we can find positive constants $A$ and $B$ such that
\begin{equation}\label{eq:ineq}
  \frac{g_{ij} (x, t ) v^i v^j}{f(x, t)} \le 
  (A+B \, |t|)^2 \delta _{ij} v^i v^j \, .
\end{equation}
  As $F(t) = A+B\, |t|$ satisfies the integral condition (\ref{eq:F}), this proves that
  our orthogonal splitting satisfies the metric growth condition. -- Our assumptions on
  $\gamma$ guarantee that we can find a curve $\gamma ' : {\mathbb{R}} \longrightarrow
  M_+$ which, in terms of our orthogonal splitting, is of the form $\gamma '( \tau ) =
  \big( \beta ' (\tau), \tau \big)$ such that $\gamma ' (\tau ) = \gamma (\tau )$ for
  all $\, ]-\infty, \, \tau _b \, ]\,$, with some $\tau _b \in {\mathbb{R}}$. (Introducing
  $\gamma '$ is necessary because $\gamma$ need not be defined on all of $\mathbb{R}$.)
  As $\gamma$ does not meet the caustic of the past light-cone of $p$, we may 
  assure that $\gamma '$ does not meet the caustic of the past light-cone of $p$.  
  As $\gamma$ does not go to the horizon or to infinity for $\tau \rightarrow - 
  \infty$, the set $\{ \, \beta ' (\tau) \, | \, -\infty < \tau < \tau_b\}$ is
  confined to a compact region. Hence, for every sequence $(\tau_i)_{i \in {\mathbb{N}}}$ 
  with $\tau_i \rightarrow - \infty$ the sequence $\big( \beta ' (\tau_i) \big) 
  _{i \in {\mathbb{N}}}$ must have a convergent subsequence. This shows that all 
  the assumptions of Theorem \ref{theo:Uh} are satisfied if we replace $\gamma$ with 
  $\gamma '$. Hence, the theorem tells us that $N_k' \ge B_k$, where $N_k'$ is the 
  number of past-pointing lightlike geodesics with index $k$ from $p$ to $\gamma '$
  and $B_k$ is the $k$-th Betti number of the loop space of $M_+ \simeq S^2 \times
  {\mathbb{R}}^2$.  As $M_+ \simeq S^2 \times {\mathbb{R}} ^2$ is simply connected but 
  not contractible to a point, the theorem of Serre \cite{Serre1951} guarantees that $B_k >0$ 
  and, thus, $N_k' >0$ for all but finitely many $k \in {\mathbb{N}}$. Hence, for 
  almost all positive integers $k$ there is a past-pointing lightlike geodesic of 
  index $k$ from $p$ to $\gamma '$. This gives us an infinite sequence $(\lambda _n)
  _{n \in {\mathbb{N}}}$ of mutually different past-pointing lightlike geodesics 
  from $p$ to $\gamma '$ such that the index of $\lambda _n$ goes to infinity if
  $n \rightarrow \infty$. We denote the point where $\lambda _n$ meets the curve 
  $\gamma '$ by $\gamma '(\tau _n )$. What remains to be shown is that $\tau _n 
  \rightarrow - \infty$ for $n \rightarrow \infty \,$; as $\gamma$ coincides with
  $\gamma '$ on $\;]-\infty, \, \tau _b \, ]$, this would make sure that all but
  finitely many $\lambda _n$ arrive indeed at $\gamma$. So we have to prove
  that it is impossible to select infinitely many $\tau _n$ that
  are bounded below. By contradiction, assume that we can find a common lower bound 
  for infinitely many $\tau _n$. As the $\tau_n$ are obviously bounded above by the 
  value of the Boyer-Lindquist time coordinate at $p$, this implies that the $\tau _n$ 
  have an accumulation point. Hence, for an infinite subsequence of our lightlike 
  geodesics $\lambda _n$ the end-points $\gamma ' (\tau _n )$ converge to some point 
  $q$ on $\gamma '$. As $\gamma '$ does not meet the caustic of the past light-cone 
  of $p$, the past light-cone of $p$ is an immersed 3-dimensional lightlike 
  submanifold near  $q$. We have thus found an infinite sequence of points $\gamma ' 
  (\tau _n )$ that lie in a 3-dimensional lightlike submanifold and, at the same 
  time, on a timelike curve. Such a sequence can converge to $q$ only if all but 
  finitely many $\gamma ' (\tau _n)$ are equal to $q$. So there are infinitely many 
  $\lambda _n$ that terminate at $q$. As there is only one lightlike direction 
  tangent to the past light-cone of $p$ at $q$, all these infinitely many lightlike 
  geodesics must have the same tangent direction at $q$. As there are no periodic 
  lightlike geodesics in the globally hyperbolic spacetime $(M_+,g)$, any two 
  lightlike geodesics from $p$ to $q$ with a common tangent direction at $q$ must 
  coincide. This contradicts the fact that the $\lambda _n$ are mutually 
  different, so our assumption that there is a common lower bound for infinitely 
  many $\tau _n$ cannot be true. 
\end{proof}

  The proof shows that in Proposition \ref{prop:infinite} the condition of $\gamma 
  (\tau )$ going neither to infinity nor to the horizon for $\tau \to - \infty$ 
  can be a little bit relaxed. It suffices to require that there is a sequence  
  $(\tau _i)_{i \in \mathbb{N}}$ of time parameters with $\tau _i \to - \infty$ 
  for $i \to \infty$ such that the spatial coordinates of $\gamma ( \tau _i )$ 
  converge. This condition is mathematically weaker than the one given in the 
  proposition, but there are probably no physically interesting situations where 
  the former is satisfied and the latter is not. 

  Proposition \ref{prop:infinite} tells us that a Kerr-Newman black hole produces infinitely
  many images for an arbitrary observer, provided that the worldline of the light source 
  satisfies some (mild) conditions. At the same time, this proposition demonstrates that 
  the past light-cone of every point $p$ in the exterior Kerr-Newman spacetime must have a non-empty
  and, indeed, rather complicated caustic; otherwise it would not be possible to find
  a sequence of past-pointing lightlike geodesics $\lambda _n$ from $p$ that intersect
  this caustic arbitrarily often for $n$ sufficiently large. Please note that the 
  last sentence of Proposition \ref{prop:infinite} makes clear that for the existence 
  of infinitely many images it is essential to assume that the light source exists 
  since arbitrarily early times. 

  In Proposition \ref{prop:shell} we have shown that all lightlike geodesics from $p$
  to $\gamma$ are confined to a spherical shell that contains the photon region $K$. 
  We can now show that, under the assumptions of Proposition \ref{prop:infinite}, 
  almost all past-pointing lightlike geodesics from $p$ to $\gamma$ come actually 
  arbitrarily close to $K$.

\begin{proposition}\label{prop:limit}
  Let $U$ be any open subset of $M_+$ that contains the region $K$ defined by
  {\em (\ref{eq:K})}.
  Then, if the assumptions of Proposition $\ref{prop:infinite}$ are
  satisfied, all but finitely many past-pointing lightlike geodesics from $p$
  to $\gamma$ intersect $U$. 
\end{proposition}
\begin{proof}
  The sequence $( \lambda _n )_{n \in {\mathbb{N}}}$ of Proposition
  \ref{prop:infinite} gives us a sequence $( w_n )_{n \in {\mathbb{N}}}$ of 
  mutually different lightlike vectors $w_n \in T_p M_+$ with $dt(w_n) = -1$
  and a sequence  $( s_n )_{n \in {\mathbb{N}}}$ of real numbers $s_n \ge 0$
  such that ${\mathrm{exp}} _p (s_n w_n)$ is on $\gamma$ for all 
  $n \in {\mathbb{N}}$. Here ${\mathrm{exp}} _p$ denotes the 
  exponential map of the Levi-Civita derivative of the Kerr-Newman metric at the point $p$.
  Since the 2-sphere consisting of the lightlike vectors $w \in T_p M_+$ with $dt(w) = -1$ 
  (which may be regarded as the observer's celestial sphere) is compact, a subsequence 
  of $( w_n )_{n \in {\mathbb{N}}}$ must converge to some lightlike vector 
  $w _{\infty} \in T_p M_+$. By Proposition \ref{prop:infinite}, 
  the sequence $\big( {\mathrm{exp}} _p (s_n w_n) \big)_{n \in {\mathbb{N}}}$ cannot 
  have an accumulation point, hence $s_n \to \infty$ for $n \to \infty$. Owing to
  Proposition \ref{prop:shell}, the radius coordinate $r$ of all points 
  ${\mathrm{exp}}_p (s w_n)$ with $s \in [0,s_n]$ is bounded, so the past-pointing 
  past-inextendible lightlike geodesic 
\begin{equation}\label{eq:exp}
\begin{split}
  \lambda _{\infty}: [0, \infty \, [ \; & \longrightarrow \, M_+ 
\\
  s \,  \longmapsto & \; \lambda _{\infty} (s) = {\mathrm{exp}}_p(sw_{\infty})
\end{split}
\end{equation}
  cannot go to infinity. Let us assume that $\lambda _{\infty}$ goes to the horizon. By
  Proposition \ref{prop:shell}, this is possible only in the extreme case $a^2 + q^2=m^2$. 
  Then along $\lambda _n$ the radius coordinate $r$ must have local minima 
  arbitrarily close to $r_+$ for $n$ sufficiently large. As, by Proposition \ref{prop:convex},
  such minima cannot lie in $M_{\mathrm{in}}$, the geodesic $\lambda _n$ has to meet $K$ 
  for $n$ sufficiently large and we are done. Therefore, we may assume for the
  rest of the proof that $\lambda _{\infty}$ does not go to the horizon. So along 
  $\lambda _{\infty}$ the coordinate $r$ must either approach a limit value 
  $r_{\infty}$ or pass through a maximum and a minimum. In the first case, 
  both the first and the second derivative of $s \longmapsto r \big( \lambda _{\infty} 
  (s) \big)$ must go to zero for $s \to \infty$. This is possible only if 
  $\lambda _{\infty}$ comes arbitrarily close to $K$, because, as we know from
  the proof of Proposition \ref{prop:convex}, the implication (\ref{eq:Xpos}) holds
  on $M_{\mathrm{out}}$ and the implication (\ref{eq:Xneg}) holds on $M_{\mathrm{in}}$. 
  In the second case, again by Proposition \ref{prop:convex},
  the maximum cannot lie in $M_{\mathrm{out}}$ and the minimum cannot lie in
  $M_{\mathrm{in}}$; hence, both the maximum and the minimum must lie in $K$.
  In both cases we have, thus, found that $\lambda _{\infty}$ and hence all but finitely 
  many $\lambda _n$ intersect $U$.
\end{proof}


\section{Discussion and concluding remarks}\label{sec:conclusion}

  We have proven, with the help of Morse theory, in Proposition \ref{prop:infinite} 
  that a Kerr-Newman black hole acts as a gravitational lens that produces 
  infinitely many images. We emphasize that we made only very mild assumptions 
  on the motion of the light source and that we considered the whole domain of 
  outer communication, including the ergosphere. For the sake of comparison, 
  the reader may consult Section 7.2 of Masiello \cite{Masiello1994} where it is shown, with
  the help of Morse theory, that a Kerr black hole produces infinitely many images. 
  However, Masiello's work is based on a special version of Morse theory which applies 
  to stationary spacetimes only; therefore he had to exclude the ergosphere from the 
  discussion, he had to require that the worldline of the light source is an 
  integral curve of the Killing vector field $\partial _t$, and he had to restrict
  to the case of slowly rotating Kerr black holes, $0 \le a^2 < a_0 ^2$ 
  with some $a_0$ that remained unspecified, instead of the whole 
  range $0 \le a^2 \le m^2$. On the basis of our Proposition \ref{prop:shell} one
  can show that Masiello's $a_0$ is equal to $m/\sqrt{2}$; this is the value 
  of $a$ where the photon region $K$ reaches the ergosphere (see Figure
  \ref{fig:K}), i.e. where $r_+^{\mathrm{ph}} = 2m$. For a Kerr spacetime with
  $m \ge a \ge m/\sqrt{2}$ we can find an event $p$ and a $t$-line in 
  $M_+ \setminus \{\mathrm{ergosphere} \}$ that can be connected by only 
  finitely many lightlike geodesics in $M_+ \setminus \{\mathrm{ergosphere} \}$.  

  If an observer sees infinitely many images of a light source, they must have
  at least one accumulation point on the observer's celestial sphere. This follows 
  immediately from the compactness of the 2-sphere. This accumulation point 
  corresponds to a limit light ray $\lambda _{\infty}$. In the proof of 
  Proposition \ref{prop:limit} we have demonstrated that $\lambda _{\infty}$ 
  comes arbitrarily close to the photon region $K$ and that either $\lambda _{\infty}$
  approaches a sphere $r=\mathrm{constant}$ or the radius coordinate along 
  $\lambda _{\infty}$ has a minimum and 
  a maximum in $K$. (In the extreme case $a^2+q^2=m^2$ the ray $\lambda _{\infty}$ 
  may go to the inner boundary of $M_+$.) This is all one can show with the help of
  Morse theory and the qualitative methods based on the sign of centrifugal-plus-Coriolis
  force. Stronger results are possible if one uses the explicit first-order form of 
  the lightlike geodesic equation in the Kerr-Newman spacetime, making use of the 
  constants of motion which reflect complete integrability. Then one can show that 
  along a lightlike geodesic in $M_+$ the radius coordinate is either monotonous
  or has precisely one turning point. (This result can be deduced, e.g., from 
  Calvani and Turolla \cite{CalvaniTurolla1981}).
  Thus, the case that there is a minimum and a maximum in $K$ is, actually, impossible.
  As a consequence, the limit light ray $\lambda _{\infty}$ necessarily approaches a 
  sphere $r=\mathrm{constant}$. By total integrability it must then approach a 
  lightlike geodesic with the same constants of motion. Of course, this must be one
  of the spherical geodesics in $K$. (In the extreme case
  $a^2+q^2=m^2$ the limit ray $\lambda _{\infty}$ may approach the 
  circular light ray at $r_+^{\mathrm{ph}}=m$ which is outside of $M_+$.)  

  Also, it follows from Proposition \ref{prop:infinite} that the limit curve
  $\lambda _{\infty}$ meets the caustic of the past light cone of $p$ infinitely
  many times. This gives, implicitly, some information on the structure of the 
  caustic. For the Kerr case, $q=0$, it was shown numerically by Rauch and 
  Blandford \cite{RauchBlandford1994} that the caustic consists of infinitely 
  many tubes with astroid cross sections. This result was supported by recent 
  analytical results by Bozza, de Luca, Scarpetta, and Sereno 
  \cite{BozzadeLucaScarpettaSereno2005}.  

  We have shown, in Proposition \ref{prop:shell}, that all 
  lightlike geodesics connecting an event $p$ to a timelike curve $\gamma$
  in the exterior Kerr-Newman spacetime $M_+$ are confined to the smallest 
  spherical shell that contains $p$, $\gamma$ and the photon region $K$. 
  If $\gamma$ satisfies the assumptions of Proposition \ref{prop:infinite}, 
  which guarantees infinitely many past-pointing lightlike geodesics from $p$ to 
  $\gamma$, Proposition \ref{prop:limit} tells us that all but finitely
  many of them come arbitrarily close to the photon region $K$. Thus, our result 
  that a Kerr-Newman black hole produces infinitely many images is crucially 
  related to the existence of the photon region. If we restrict to some open subset 
  of $M_+$ whose closure is completely contained in either $M_{\mathrm{out}}$ or 
  $M_{\mathrm{in}}$, then we are left with finitely many images for any choice of 
  $p$ and $\gamma$. In Section \ref{sec:centrifugal} we have seen that the 
  decomposition of $M_+$ into $M_{\mathrm{in}}$, $M_{\mathrm{out}}$ and 
  the photon region $K$ plays an important role in view of 
  centrifugal-plus-Coriolis force reversal; if we restrict to an open subset 
  of $M_+$ that is contained in either $M_{\mathrm{out}}$ or $M_{\mathrm{in}}$,
  then we are left with a spacetime on which $\partial _r \Psi_+ $ and 
  $\partial _r \Psi_- $ have the same sign, i.e., the centrifugal-plus-Coriolis
  force for large velocities points either always outwards or always inwards. In 
  an earlier paper \cite{HassePerlick2002} we have shown that in a 
  spherically symmetric and static spacetime the occurrence of gravitational 
  lensing with infinitely many images is equivalent to the occurrence of 
  centrifugal force reversal. Our new results demonstrate that the same 
  equivalence is true for subsets of the exterior Kerr-Newman spacetime, 
  with the only difference that instead of the centrifugal force alone 
  now we have to consider the sum of centrifugal and Coriolis force in the limit 
  $v \to 1$. It is an interesting problem to inquire whether this observation carries 
  over to other spacetimes with two commuting Killing vector fields $\partial _t$ 
  and $\partial _{\varphi}$ that span timelike 2-surfaces with cylindrical topology.


\section*{Acknowledgment}

V. P. wishes to thank Simonetta Frittelli and Arlie Petters for inviting him to the 
workshop on ``Gravitational lensing in the Kerr spacetime geometry'' at the American Institute 
of Mathematics, Palo Alto, July 2005, and all participants of this workshop for stimulating
and useful discussions.



\begin{thebibliography}{29}
\expandafter\ifx\csname natexlab\endcsname\relax\def\natexlab#1{#1}\fi
\expandafter\ifx\csname bibnamefont\endcsname\relax
  \def\bibnamefont#1{#1}\fi
\expandafter\ifx\csname bibfnamefont\endcsname\relax
  \def\bibfnamefont#1{#1}\fi
\expandafter\ifx\csname citenamefont\endcsname\relax
  \def\citenamefont#1{#1}\fi
\expandafter\ifx\csname url\endcsname\relax
  \def\url#1{\texttt{#1}}\fi
\expandafter\ifx\csname urlprefix\endcsname\relax\def\urlprefix{URL }\fi
\providecommand{\bibinfo}[2]{#2}
\providecommand{\eprint}[2][]{\url{#2}}

\bibitem[{\citenamefont{Darwin}(1959)}]{Darwin1959}
\bibinfo{author}{\bibfnamefont{C.}~\bibnamefont{Darwin}},
  \bibinfo{journal}{Proc. Roy. Soc. London} \textbf{\bibinfo{volume}{A 249}},
  \bibinfo{pages}{180} (\bibinfo{year}{1959}).

\bibitem[{\citenamefont{Uhlenbeck}(1975)}]{Uhlenbeck1975}
\bibinfo{author}{\bibfnamefont{K.}~\bibnamefont{Uhlenbeck}},
  \bibinfo{journal}{Topology} \textbf{\bibinfo{volume}{14}},
  \bibinfo{pages}{69} (\bibinfo{year}{1975}).

\bibitem[{\citenamefont{Abramowicz et~al.}(1988)\citenamefont{Abramowicz,
  Carter, and Lasota}}]{AbramowiczCarterLasota1988}
\bibinfo{author}{\bibfnamefont{M.~A.} \bibnamefont{Abramowicz}},
  \bibinfo{author}{\bibfnamefont{B.}~\bibnamefont{Carter}}, \bibnamefont{and}
  \bibinfo{author}{\bibfnamefont{J.-P.} \bibnamefont{Lasota}},
  \bibinfo{journal}{Gen. Relativ. Gravit.} \textbf{\bibinfo{volume}{20}},
  \bibinfo{pages}{1173} (\bibinfo{year}{1988}).

\bibitem[{\citenamefont{Abramowicz}(1990)}]{Abramowicz1990}
\bibinfo{author}{\bibfnamefont{M.~A.} \bibnamefont{Abramowicz}},
  \bibinfo{journal}{Mon. Not. Roy. Astron. Soc.}
  \textbf{\bibinfo{volume}{245}}, \bibinfo{pages}{733} (\bibinfo{year}{1990}).

\bibitem[{\citenamefont{Abramowicz et~al.}(1993)\citenamefont{Abramowicz,
  Nurowski, and Wex}}]{AbramowiczNurowskiWex1993}
\bibinfo{author}{\bibfnamefont{M.~A.} \bibnamefont{Abramowicz}},
  \bibinfo{author}{\bibfnamefont{P.}~\bibnamefont{Nurowski}}, \bibnamefont{and}
  \bibinfo{author}{\bibfnamefont{N.}~\bibnamefont{Wex}},
  \bibinfo{journal}{Class. Quant. Grav.} \textbf{\bibinfo{volume}{10}},
  \bibinfo{pages}{L183} (\bibinfo{year}{1993}).

\bibitem[{\citenamefont{McKenzie}(1985)}]{McKenzie1985}
\bibinfo{author}{\bibfnamefont{R.}~\bibnamefont{McKenzie}},
  \bibinfo{journal}{J. Math. Phys.} \textbf{\bibinfo{volume}{26}},
  \bibinfo{pages}{1592} (\bibinfo{year}{1985}).

\bibitem[{\citenamefont{Giannoni and Masiello}(1996)}]{GiannoniMasiello1996}
\bibinfo{author}{\bibfnamefont{F.}~\bibnamefont{Giannoni}} \bibnamefont{and}
  \bibinfo{author}{\bibfnamefont{A.}~\bibnamefont{Masiello}},
  \bibinfo{journal}{Gen. Relativ. Gravit.} \textbf{\bibinfo{volume}{28}},
  \bibinfo{pages}{855} (\bibinfo{year}{1996}).

\bibitem[{\citenamefont{Giannoni et~al.}(1998)\citenamefont{Giannoni, Masiello,
  and Piccione}}]{GiannoniMasielloPiccione1998}
\bibinfo{author}{\bibfnamefont{F.}~\bibnamefont{Giannoni}},
  \bibinfo{author}{\bibfnamefont{A.}~\bibnamefont{Masiello}}, \bibnamefont{and}
  \bibinfo{author}{\bibfnamefont{P.}~\bibnamefont{Piccione}},
  \bibinfo{journal}{Ann. Inst. H. Poincar{\'e}} \textbf{\bibinfo{volume}{69}},
  \bibinfo{pages}{359} (\bibinfo{year}{1998}).

\bibitem[{\citenamefont{Kovner}(1990)}]{Kovner1990}
\bibinfo{author}{\bibfnamefont{I.}~\bibnamefont{Kovner}},
  \bibinfo{journal}{Astrophys. J.} \textbf{\bibinfo{volume}{351}},
  \bibinfo{pages}{114} (\bibinfo{year}{1990}).

\bibitem[{\citenamefont{Perlick}(1990)}]{Perlick1990b}
\bibinfo{author}{\bibfnamefont{V.}~\bibnamefont{Perlick}},
  \bibinfo{journal}{Class. Quant. Grav.} \textbf{\bibinfo{volume}{7}},
  \bibinfo{pages}{1319} (\bibinfo{year}{1990}).

\bibitem[{\citenamefont{Hawking and Ellis}(1973)}]{HawkingEllis1973}
\bibinfo{author}{\bibfnamefont{S.~W.} \bibnamefont{Hawking}} \bibnamefont{and}
  \bibinfo{author}{\bibfnamefont{G.~F.~R.} \bibnamefont{Ellis}},
  \emph{\bibinfo{title}{The large scale structure of space-time}}
  (\bibinfo{publisher}{Cambridge University Press},
  \bibinfo{address}{Cambridge}, \bibinfo{year}{1973}).

\bibitem[{\citenamefont{Bernal and S{\'a}nchez}(2005)}]{BernalSanchez2005}
\bibinfo{author}{\bibfnamefont{A.}~\bibnamefont{Bernal}} \bibnamefont{and}
  \bibinfo{author}{\bibfnamefont{M.}~\bibnamefont{S{\'a}nchez}},
  \bibinfo{journal}{Comm. Math. Phys.} \textbf{\bibinfo{volume}{257}},
  \bibinfo{pages}{43} (\bibinfo{year}{2005}).

\bibitem[{\citenamefont{Perlick}(2000)}]{Perlick2000}
\bibinfo{author}{\bibfnamefont{V.}~\bibnamefont{Perlick}}, in
  \emph{\bibinfo{booktitle}{Einstein's Field Equations and their Physical
  Implications: Selected Essays in Honour of J{\"u}rgen Ehlers}}, edited by
  \bibinfo{editor}{\bibfnamefont{B.}~\bibnamefont{Schmidt}}
  (\bibinfo{publisher}{Springer}, \bibinfo{address}{Berlin, Germany},
  \bibinfo{year}{2000}), vol. \bibinfo{volume}{540} of
  \emph{\bibinfo{series}{Lecture Notes in Physics}}, pp.
  \bibinfo{pages}{373--425}.

\bibitem[{\citenamefont{Frankel}(1997)}]{Frankel1997}
\bibinfo{author}{\bibfnamefont{T.}~\bibnamefont{Frankel}},
  \emph{\bibinfo{title}{The geometry of physics}}
  (\bibinfo{publisher}{Cambridge University Press},
  \bibinfo{address}{Cambridge}, \bibinfo{year}{1997}).

\bibitem[{\citenamefont{Serre}(1951)}]{Serre1951}
\bibinfo{author}{\bibfnamefont{J.~P.} \bibnamefont{Serre}},
  \bibinfo{journal}{Ann. Math.} \textbf{\bibinfo{volume}{54}},
  \bibinfo{pages}{425} (\bibinfo{year}{1951}).

\bibitem[{\citenamefont{Misner et~al.}(1973)\citenamefont{Misner, Thorne, and
  Wheeler}}]{MisnerThorneWheeler1973}
\bibinfo{author}{\bibfnamefont{C.}~\bibnamefont{Misner}},
  \bibinfo{author}{\bibfnamefont{K.}~\bibnamefont{Thorne}}, \bibnamefont{and}
  \bibinfo{author}{\bibfnamefont{J.~A.} \bibnamefont{Wheeler}},
  \emph{\bibinfo{title}{Gravitation}} (\bibinfo{publisher}{Freeman},
  \bibinfo{address}{San Francisco}, \bibinfo{year}{1973}).

\bibitem[{\citenamefont{Chandrasekhar}(1983)}]{Chandrasekhar1983}
\bibinfo{author}{\bibfnamefont{S.}~\bibnamefont{Chandrasekhar}},
  \emph{\bibinfo{title}{The mathematical theory of black holes}}
  (\bibinfo{publisher}{Oxford University Press}, \bibinfo{address}{Oxford},
  \bibinfo{year}{1983}).

\bibitem[{\citenamefont{O'Neill}(1995)}]{ONeill1995}
\bibinfo{author}{\bibfnamefont{B.}~\bibnamefont{O'Neill}},
  \emph{\bibinfo{title}{The geometry of {K}err black holes}}
  (\bibinfo{publisher}{A. K. Peters}, \bibinfo{address}{Wellesley},
  \bibinfo{year}{1995}).

\bibitem[{\citenamefont{Foertsch et~al.}(2003)\citenamefont{Foertsch, Hasse,
  and Perlick}}]{FoertschHassePerlick2003}
\bibinfo{author}{\bibfnamefont{T.}~\bibnamefont{Foertsch}},
  \bibinfo{author}{\bibfnamefont{W.}~\bibnamefont{Hasse}}, \bibnamefont{and}
  \bibinfo{author}{\bibfnamefont{V.}~\bibnamefont{Perlick}},
  \bibinfo{journal}{Class. Quant. Grav.} \textbf{\bibinfo{volume}{20}},
  \bibinfo{pages}{4635} (\bibinfo{year}{2003}).

\bibitem[{\citenamefont{Nayak and Vishveshwara}(1996)}]{NayakVishveshwara1996}
\bibinfo{author}{\bibfnamefont{K.~R.} \bibnamefont{Nayak}} \bibnamefont{and}
  \bibinfo{author}{\bibfnamefont{C.~V.} \bibnamefont{Vishveshwara}},
  \bibinfo{journal}{Class. Quant. Grav.} \textbf{\bibinfo{volume}{13}},
  \bibinfo{pages}{1783} (\bibinfo{year}{1996}).

\bibitem[{\citenamefont{de~Felice and
  Usseglio-Tomasset}(1991)}]{deFeliceUsseglio1991}
\bibinfo{author}{\bibfnamefont{F.}~\bibnamefont{de~Felice}} \bibnamefont{and}
  \bibinfo{author}{\bibfnamefont{S.}~\bibnamefont{Usseglio-Tomasset}},
  \bibinfo{journal}{Class. Quant. Grav.} \textbf{\bibinfo{volume}{8}},
  \bibinfo{pages}{1871} (\bibinfo{year}{1991}).

\bibitem[{\citenamefont{Kozlowski et~al.}(1978)\citenamefont{Kozlowski,
  Jaroszynski, and Abramowicz}}]{KozlowskiJaroszynskiAbramowicz1978}
\bibinfo{author}{\bibfnamefont{M.}~\bibnamefont{Kozlowski}},
  \bibinfo{author}{\bibfnamefont{M.}~\bibnamefont{Jaroszynski}},
  \bibnamefont{and} \bibinfo{author}{\bibfnamefont{M.~A.}
  \bibnamefont{Abramowicz}}, \bibinfo{journal}{Astron. \& Astrophys.}
  \textbf{\bibinfo{volume}{63}}, \bibinfo{pages}{209} (\bibinfo{year}{1978}).

\bibitem[{\citenamefont{Stuchlik et~al.}(2000)\citenamefont{Stuchlik, Hledik,
  and Jur{\'a}n}}]{StuchlikHledikJuran2000}
\bibinfo{author}{\bibfnamefont{Z.}~\bibnamefont{Stuchlik}},
  \bibinfo{author}{\bibfnamefont{S.}~\bibnamefont{Hledik}}, \bibnamefont{and}
  \bibinfo{author}{\bibfnamefont{J.}~\bibnamefont{Jur{\'a}n}},
  \bibinfo{journal}{Class. Quant. Grav.} \textbf{\bibinfo{volume}{17}},
  \bibinfo{pages}{2691} (\bibinfo{year}{2000}).

\bibitem[{\citenamefont{Masiello}(1994)}]{Masiello1994}
\bibinfo{author}{\bibfnamefont{A.}~\bibnamefont{Masiello}},
  \emph{\bibinfo{title}{Variational methods in {L}orentzian geometry}}, Pitman
  Research Notes in Mathematics Series 309 (\bibinfo{publisher}{Longman
  Scientific \& Technical}, \bibinfo{address}{Essex}, \bibinfo{year}{1994}).

\bibitem[{\citenamefont{Calvani and Turolla}(1981)}]{CalvaniTurolla1981}
\bibinfo{author}{\bibfnamefont{M.}~\bibnamefont{Calvani}} \bibnamefont{and}
  \bibinfo{author}{\bibfnamefont{R.}~\bibnamefont{Turolla}},
  \bibinfo{journal}{J. Phys. A} \textbf{\bibinfo{volume}{14}},
  \bibinfo{pages}{1931} (\bibinfo{year}{1981}).

\bibitem[{\citenamefont{Rauch and Blandford}(1994)}]{RauchBlandford1994}
\bibinfo{author}{\bibfnamefont{K.}~\bibnamefont{Rauch}} \bibnamefont{and}
  \bibinfo{author}{\bibfnamefont{R.}~\bibnamefont{Blandford}},
  \bibinfo{journal}{Astrophys. J.} \textbf{\bibinfo{volume}{421}},
  \bibinfo{pages}{46} (\bibinfo{year}{1994}).

\bibitem[{\citenamefont{Bozza et~al.}(2005)\citenamefont{Bozza, de~Luca,
  Scarpetta, and Sereno}}]{BozzadeLucaScarpettaSereno2005}
\bibinfo{author}{\bibfnamefont{V.}~\bibnamefont{Bozza}},
  \bibinfo{author}{\bibfnamefont{F.}~\bibnamefont{de~Luca}},
  \bibinfo{author}{\bibfnamefont{G.}~\bibnamefont{Scarpetta}},
  \bibnamefont{and} \bibinfo{author}{\bibfnamefont{M.}~\bibnamefont{Sereno}},
  \bibinfo{journal}{Phys. Rev. D} \textbf{\bibinfo{volume}{72}},
  \bibinfo{pages}{083003} (\bibinfo{year}{2005}).

\bibitem[{\citenamefont{Hasse and Perlick}(2002)}]{HassePerlick2002}
\bibinfo{author}{\bibfnamefont{W.}~\bibnamefont{Hasse}} \bibnamefont{and}
  \bibinfo{author}{\bibfnamefont{V.}~\bibnamefont{Perlick}},
  \bibinfo{journal}{Gen. Relativ. Gravit.} \textbf{\bibinfo{volume}{34}},
  \bibinfo{pages}{415} (\bibinfo{year}{2002}).

\bibitem[{\citenamefont{Perlick}(2004)}]{Perlick2004}
\bibinfo{author}{\bibfnamefont{V.}~\bibnamefont{Perlick}},
  \bibinfo{journal}{Living Rev. Relativity} \textbf{\bibinfo{volume}{7(9)}}
  (\bibinfo{year}{2004}),
  \bibinfo{note}{http://www.livingreviews.org/lrr-2004-9}.

\end{thebibliography}


\newpage

\newpage

\end{document}